\documentclass[preprintnumbers,amsmath,amssymb,nofootinbib]{revtex4}

\usepackage{color}
\usepackage{graphicx}
\usepackage{dcolumn}
\usepackage{bm}
\usepackage{latexsym}
\usepackage{epsfig}
\usepackage{rotating}
\usepackage{bm}

\newcommand{\be}{\begin{equation}}
\newcommand{\ee}{\end{equation}}
\newcommand{\beqq}{\setlength\arraycolsep{2pt}\begin{eqnarray}}
\newcommand{\eeqq}{\vspace{0cm} \end{eqnarray}}
\newcommand{\bea}{\begin{eqnarray}}
\newcommand{\eea}{\end{eqnarray}}

\begin{document}

\title{An unified cosmological model driven by a scalar field\\ nonminimally coupled to gravity}

\author{S. H. Pereira$^{}$}\email{s.pereira@unesp.br}

\affiliation{$^{}$Universidade Estadual Paulista (Unesp)\\Faculdade de Engenharia de Guaratinguet\'a \\ Departamento de F\'isica\\ Av. Dr. Ariberto Pereira da Cunha 333\\
12516-410 -- Guaratinguet\'a, SP, Brazil}




\begin{abstract}

This paper consider an universe dominated by baryonic matter, radiation and a nonminimally coupled massive scalar field under the action of a symmetry breaking potential. Inflation occurs naturally with appropriated slow-roll values. The field evolves up to late times, when it is supposed to enter a slowing varying phase, remaining at rest at the value $\phi_0$. A mechanism for coupling the scalar field to baryonic and radiation energy densities through the nonminimal coupling $\xi$ is presented, leading naturally to dark matter and dark radiation components. A cosmological constant like term is also present, acting as a dark energy at late times. All the density parameters are in good agreement to observational data of Hubble measurements plus Supernovae Type Ia data. All the desired phases of cosmic evolution appears naturally in this simple model. 

\end{abstract}

\maketitle

\section{Introduction}
Unifying all phases of cosmic evolution is an old task for physicists. Inflation, radiation, dark matter (DM) and dark energy (DE) represent almost the whole history of the Universe \cite{kolb,bookliddle,cosmology}, each one endowed with specific characteristics which may be confronted to observations. A dark radiation (DR) content has also been proposed recently as responsible to mediates interactions of dark matter \cite{darkradiation}.


The simplest unifying models are based on a single scalar field with a potential $V(\phi)=V_0+\frac{1}{2}m^2\phi^2$ and some additional suppositions, as the impositions needed to correctly describe the reheating phase after inflation. The presence of a small $V_0$ constant is needed to explain the present dark energy dominance. The main features of such kind of model were discussed by Liddle and Ure\~na-L\'opez \cite{liddle2006,liddle2008}, which showed that an incomplete decay of the inflaton field would be suffice to allow such unification. The model was implemented and discussed again later in different contexts \cite{cardenas2007,pano2007,hidalgo2012,bastero2016}. Linde \cite{linde2002} has also showed that the mass of the scalar field in order to obtain the correct amplitude of primordial scalar perturbations in usual chaotic inflation scenario is $m\simeq 10^{-6}m_{pl}$. Also, in order to satisfies the measured value of the present dark matter mass per photon from WMAP, the important constraint $(m/m_{pl})^{1/2}\phi_*^2/m_{pl}^2 \simeq 4\times 10^{-29}$ must be satisfied \cite{liddle2006}, where $\phi_*$ is the initial scalar amplitude at the time $t_*$ for which the scalar mass $m$ equals the Hubble parameter $H$, namely $m=H_*$. The above relation will be called DM/photon ratio, for future reference. After a standard inflation driven by the scalar field during $H \gg m$, the scalar energy density evolves as $\rho_\phi=\frac{1}{2}m^2\phi_*^2a_*^3a^{-3}$ for $t>t_*$ and $H<m$, where $a$ is the scale factor, with $\phi$ oscillating with amplitudes smaller than $\phi_*$ and justifying the necessity of an incomplete decay of the field in order to follow a dark matter evolution as $a^{-3}$. For late times the constant potential $V_0$ dominates and dark energy evolution takes place. These are the main ingredients of the triple unification proposed by \cite{liddle2006,liddle2008}, namely inflation, dark matter and dark energy. A triple unification in a two-scalar-field cosmological model was recently studied in \cite{Paulo2020}. Finally, is also important to cite an inflationary and dark energy unifying model \cite{esha2015} based on a scalar field with nonminimal coupling $\xi$ to gravity.   

In the present paper it is proposed a new model to unify inflation, dark matter, dark energy and also furnish a kind of dark radiation contribution to the universe content. The model is endowed with a quadratic mass term plus a symmetry broken term nonminimally coupled to gravity.
The direct coupling of the scalar field to standard baryonic matter and also to radiation will furnish the dark matter and dark radiation like components, while the quadratic mass term will be responsible to a dark energy accelerated phase, similar to an effective cosmological constant term. The mechanism to couple the scalar field to baryons and radiation is presented in details for the case of a slowly varying scalar field in a kind of `toy model'. The nonminimal coupling of the scalar field with the gravitational field act as an amplifier of the baryon and radiation components, given rise to the effects of dark matter and dark radiation in a natural way.

\section{Dynamic equations}

 The action for a real scalar field nonminimally coupled to gravity is given by:
\begin{equation}
S = \int d^4 x \sqrt{-g} \left[ \frac{{R}}{2\kappa^2} -\frac{\xi}{2}R\phi^2 + \mathcal{L}_\phi +\mathcal{L}_m\right] \,,
\label{actionE}
\end{equation}
where $\kappa^2\equiv8\pi G = \frac{8\pi}{m_{pl}^{2}}$ with $c=1$, $\xi$ is the nonminimal coupling parameter between curvature scalar $R$ and the scalar field and $\mathcal{L}_m$ stands for the Lagrangian of the ordinary matter. The Lagragian for scalar field is\footnote{The metric is $-,\,+,\,+,\,+$.}:
\begin{equation}
    \mathcal{L}_\phi=-\frac{1}{2}\nabla^\mu \phi\nabla_\mu \phi -V(\phi)\,.
\end{equation}

In a flat Friedmann-Robertson-Walker (FRW) background, the Friedmann equations and the dynamic field equation for a homogeneous and isotropic field $\phi(t)$ are \cite{faraoni2001}:
\begin{equation}
H^2=\frac{\kappa^2}{3}\bigg[\frac{\dot{\phi}^2}{2}+V(\phi)+3\xi (H^2 \phi^2+2 \phi\dot{\phi}H)+\rho_m\bigg]\,,\label{H2}
\end{equation}
\begin{equation}
    \dot{H}=-\frac{\kappa^2}{2}\bigg[ \dot{\phi}^2 +2\xi(H\phi\dot{\phi}-\dot{\phi}^2-\phi\ddot{\phi}-\dot{H}\phi^2)+\rho_m+p_m\bigg]\,,\label{Hdot}
\end{equation}
\begin{equation}
\ddot{\phi}+3H\dot{\phi}+V'- 6\xi\phi(\dot{H}+2H^2)=0\,.\label{eqphi}
\end{equation}
with $H=\dot{a}/a$, $V' = {dV(\phi)/ d\phi}$, $\rho_m$ and $p_m$ are the corresponding energy density and pressure for ordinary matter, respectively. Equations (\ref{H2})-(\ref{eqphi}) generalize the ones for a minimally coupled standard scalar field, with $\xi=0$.

In order to study the evolution of $a(t)$ and $\phi(t)$ we choose to use a quadratic mass term plus a symmetry breaking potential, namely:
\begin{equation}
V(\phi)=\frac{1}{2}m^2\phi^2 + A^4\Bigg(1-\frac{\phi^2}{\sigma^2}\Bigg)^2 \,,\label{V}
\end{equation}
where $m$ is the physical mass of the scalar field, $A$ is a positive constant with dimension $[M]$, which can be fixed by initial condition of inflationary phase, $\sigma$ also has dimension $[M]$, corresponding to the true vacuum of the symmetry breaking potential, which can be determined by constraints to observational data.

Such potential has a minimum at $\phi_0 = \pm \sigma\sqrt{1-\frac{m^2\sigma^2}{4A^4}}$, assuming the value:
\begin{equation}
    V_0 \equiv V(\phi_0)=\frac{1}{2}m^2\sigma^2 - \frac{m^4 \sigma^4}{16 A^4}\,. \label{V0}
\end{equation}
Notice that for a massless scalar field the minimum of the potential goes to zero, while it is positive for $\frac{m^2\sigma^2}{8A^4}<1$. At the end of evolution, in the limit $\frac{A^2}{\sigma}\gg m$ we have $\phi_0\approx \sigma$ and the first term of the potential dominates, acting as a cosomological constant term. The physical mass $m$ at this stage dominates and must be used to verify the DM/photon relation. This field also admits an effective running mass-squared $m_e^2 \equiv \frac{d^2V}{d\phi^2} = \frac{12A^4\phi^2}{\sigma^4}-\frac{4A^4}{\sigma^2}+m^2$, which is important during the final inflationary phase, when $\phi\gtrsim \sigma$, assuming a value $m^2_{e}\simeq\frac{8A^4}{\sigma^2}$.
 The value $m_{e} \approx 10^{-6}m_{pl}$ is necessary to reproduces correct density perturbations at the end of inflation in agreement with observations.

\section{Inflation}

Inflationary models with nonminimally coupled scalar fields were studied in the past \cite{futamase89,accetta85,fakir90a,fakir90b,komatsu99} and have recently returned as suitable models for describing recent CMB measurements by the Planck satellite in the Higgs inflationary scenery \cite{bezrukov2008,martin2014,takahashi2020}. These last works are based on large values of the conformal coupling constant, $\xi \gg 1$, in order to avoid the need for a fine-tuning in the field's self-coupling constant, an important ingredient to makes Higgs inflation feasible. However, Faraoni \cite{faraoni2001} has argued  that viable inflationary models with scalar fields non-minimally coupled to gravity must have $\xi=1/6$ in order to avoid a physical pathology of massive fields propagating along the light cones. Another critique to large $\xi$ values was done by \cite{barbon2009}.

In the present work it is not assumed that the scalar field is a Higgs field, thus the results will be presented for arbitrary $\xi$ following the discussions of \cite{faraoni2001}. If the inflation is driven just by a scalar field characterized by a potential $V(\phi)$ which admits a minimum $V_0$ at $\phi_0 > 0$, the condition for there exist stability around an attractor point $\phi_0$ is given by:
\begin{equation}
    V_0'' \geq \frac{V_0'}{\phi_0}\frac{1-3\xi \kappa^2 \phi_0}{1-\xi \kappa^2 \phi_0}\label{Vll}
\end{equation}
Stability also depends on the value of $\xi$, however in the particular situation that $V(\phi)$ has a minimum at $\phi_0$, satisfying $V'_0 = 0$ and $V''_0 > 0$, the $\xi$ dependence disappears and stability holds irrespective of the value of $\xi$. This is the case of the particular potential (\ref{V}), where the condition (\ref{Vll}) is satisfied if $m<\frac{2A^2}{\sigma}$, which is the same condition for $\phi_0$ be real. In this context it is possible to conclude that the potential (\ref{V}) can drive inflation in a satisfactory way.

For a more quantitative analysis on the evolution of the field $\phi$, we will divide it into three different stages, namely $\phi_i$ characterizing the beginning of inflationary phase, $\phi_f$ characterizing the end of inflation and $\phi_0$ the present day value of the scalar field. It is expected $\phi_0 \lesssim \phi_f \ll \phi_i$, with inflation occurring for $\phi_f < \phi < \phi_i$.

The Friedmann equation (\ref{H2}) can be put in the form:
\begin{equation}
H^2=\bigg(\frac{1}{1-{\kappa^2\xi \phi^2}}\bigg)\frac{\kappa^2}{3}\bigg[\frac{\dot{\phi}^2}{2}+6\xi \phi\dot{\phi}H+V(\phi) \bigg]\,,\label{H2c}
\end{equation}
The extended slow-roll parameters in order to inflation occur are given by \cite{chiba2008}:
\begin{equation}
    \epsilon = \frac{\Omega V'^2_{eff}}{2\kappa^2 V^2}\hspace{1cm} \eta = \frac{\Omega V''_{eff}}{\kappa^2 V}\hspace{1cm} \delta = \frac{\Omega' V'_{eff}}{\kappa^2 V}\,,\label{slp}
\end{equation}
where $\Omega = 1-\kappa^2\xi\phi^2$, $V'_{eff} = \frac{\Omega^2}{g}\Big(\frac{V}{\Omega^2}\Big)'$ and $g = 1 + \frac{3\Omega'^2}{2\kappa^2\Omega}$. Inflation occurs for $\epsilon \ll 1$, $|\eta|\ll 1$, $|\delta|\ll 1$ plus two additional  subsidiary conditions, namely $|\Omega\Omega''/\Omega'^2| \approx O(1)$ and $|V'_{eff}/V'| \approx O(1)$. When one of the subsidiary conditions are not satisfied the product $|\Omega\Omega''/\Omega'^2||\Omega'V'_{eff}/\kappa^2V| \ll 1$ can eventually turns the slow-roll inflation possible  (see \cite{chiba2008} for an analysis on this last condition). The inflationay phase can be studied according to different limit values of $\kappa^2\xi \phi^2$, namely $\kappa^2|\xi| \phi^2\gg 1$ or $\kappa^2|\xi| \phi^2 \ll 1$. The general potential (\ref{V}) furnish a quite complicated set of slow-roll parameters, which can be simplified with the assumption $m\ll \sqrt{2}A^2\phi/\sigma^2$, so just the second term of the potential dominates during inflation. The following three different limits will be studied in this particular case:

{\bf i} For $\kappa^2|\xi| \phi^2 \ll 1$ the slow-roll parameters are:
\begin{equation}
    \epsilon \approx \frac{m_{pl}^2\phi^2}{\pi(\phi^2-\sigma^2)^2}\hspace{1cm}  \eta \approx \frac{m_{pl}^2(3\phi^2-\sigma^2)}{2\pi(\phi^2-\sigma^2)^2}\hspace{1cm} \delta \approx -\frac{8\xi\phi^2}{\phi^2-\sigma^2}\,\label{c1}
\end{equation}
which admits inflation for $\phi \gg \sigma $ and $\phi \gg m_{pl}$. Additionally, $|\xi| \ll 1$ is required by the $\delta$ parameter. The first subsidiary condition is not satisfied since $|\Omega\Omega''/\Omega'^2| \simeq 1/2\kappa^2\xi\phi^2$, but the product $|\Omega\Omega''/\Omega'^2||\Omega'V'_{eff}/\kappa^2V| \simeq 4/\kappa^2\phi^2$ is small, thus slow-roll inflation can eventually be possible. Inflation ends when $\phi = \phi_f \approx m_{pl}$.

{\bf ii} For $\kappa^2|\xi| \phi^2 \gg 1$ the Eq. (\ref{H2c}) requires $\xi$ negative. With $|\xi| \ll 1$ the slow-roll parameters are:
\begin{equation}
    \epsilon \approx -\frac{8\sigma^4 \xi}{\phi^4}\hspace{1cm}  \eta \approx -\frac{4\sigma^2\xi}{\phi^2}\hspace{1cm} \delta \approx -\frac{8\sigma^2\xi}{\phi^2}\,\label{c2}
\end{equation}
what ensures a slow-roll inflation. However the second subsidiary condition is not satisfied but again the product $|\Omega\Omega''/\Omega'^2||\Omega'V'_{eff}/\kappa^2V| \simeq 4\sigma^2\xi/\phi^2$ is small, which shows that slow-roll inflation is also possible. Inflation ends when $\phi = \phi_f \simeq \sigma \sqrt{|\xi|}$. 

{\bf iii} For $\kappa^2|\xi| \phi^2 \gg 1$ and $|\xi|\gg 1$ the slow-roll parameters are:
\begin{equation}
    \epsilon \approx -\frac{2\sigma^4}{9\phi^4\xi}\hspace{1cm}  \eta \approx \frac{2\sigma^2}{3\phi^2}\hspace{1cm} \delta \approx \frac{4\sigma^2}{3\phi^2}\,\label{c3}
\end{equation}
and $|\Omega\Omega''/\Omega'^2||\Omega'V'_{eff}/\kappa^2V| \simeq 2\sigma^2/3\phi^2 \ll 1$. In this case inflation ends for $\phi = \phi_f \simeq \sigma$.

Notice that for the inflationary case ({\bf i}) the end of inflation is associated to a mass scale of order of $m_{pl}$, while the cases ({\bf ii}) and ({\bf iii}) are not. Other more involving cases where the mass term of the potential is relevant can also be analysed.

\section{Dark matter, dark radiation and dark energy}

Now it will presented the mechanism by which the scalar field couples to baryons and radiation, producing a dark matter component and a kind of dark radiation contribution, beyond a dark energy term coming from the minimum of the potential.

 Inflation ends after a long time of oscillations around $\phi_0$, thermalizing with radiation and baryonic matter. Thus, it is supposed the field stops to oscillate and enters a slowly varying phase satisfying $\dot{\phi}\approx 0$, which is equivalent to stay at rest in $\phi=\phi_0 = \sigma \sqrt{1-\frac{m^2\sigma^2}{4A^4}}$ with the potential given by (\ref{V}). From the three inflationary scenery analysed in last section it can be seen that inflation ends for $\phi \lesssim \sigma$, which is in agreement to the final value $\phi_0$. Such kind of incomplete decay of the field has important consequences to both dark matter and dark energy evolution driven by the scalar field, and will also be seen that a dark radiation term is also present. This corresponds to the late time evolution of the universe. Now it will be presented the mechanism that generates all these three dark sectors of the universe.

 It is supposed a universe filled with baryonic matter, radiation and the scalar field at rest in $\phi = \phi_0$. The first Friedmann equation (\ref{H2}) at this stage is:
\begin{equation}
H^2=\bigg(\frac{1}{1-\kappa^2\xi \phi_0^2} \bigg)\frac{\kappa^2}{3}\bigg[\rho_b + \rho_r + V_0\bigg]\,,\label{H2a}
\end{equation}
with $V_0$ given by (\ref{V0}). The new term inside curl-bracket is due to the nonminimal coupling of the scalar field with gravity. Defining $\alpha \equiv {\kappa^2\xi\phi_0^2}$ it is possible to write:
\begin{equation}
\frac{1}{1-\alpha}=1+\alpha+\alpha^2+\alpha^3+\dots = 1 + f(\alpha)\,,\label{f}
\end{equation}
with $f(\alpha)=\alpha+\alpha^2+\alpha^3+\dots$ an infinite series whose convergence is warranted if $|\alpha|<1$, which puts an upper limit to the final value of the field $\phi_0$, namely $\phi_0^2<\frac{m_{pl}^2}{8\pi\xi}$, which must be verified by confronting to observational data. With such a definition, equation (\ref{H2a}) can be rewritten as:
\begin{equation}
H^2=\frac{8\pi G}{3}\bigg[ \rho_b+ \rho_r + f\rho_b + f\rho_r + (1+f)V_0\bigg] \,.\label{H2f}
\end{equation}
Written in this form the term $f\rho_b = \frac{\alpha}{1-\alpha}\rho_b$ represents the gravitational coupling of the scalar field at rest, $\phi_0$, to the ordinary baryonic matter, which can be interpreted as the dark matter energy density contribution, $\rho_{dm}\equiv f\rho_b$. Notice that the factor $f$ act as an `amplifier' of the baryonic energy density. Since the evolution of the baryonic matter is $\rho_b=\rho_{b0}a^{-3}=\rho_{b0}(1+z)^3$, this also will be the evolution of the dark matter component, as occur in \cite{liddle2006}. The explicit presence of the nonminimal coupling $\xi$ and of the squared of the rest value of the field $\phi_0$ into $\alpha$ shows that the dark matter component can be interpreted as a kind of nonminimal coupling of the scalar field to the baryonic matter, acting as a `cloud' of scalar field around the standard baryonic matter. This could explain very well the presence of dark matter around galaxy clusters and even in the intergalactic medium. The same analogy can be done with the term $f\rho_r=\frac{\alpha}{1-\alpha}\rho_r$, which can be interpreted as a kind of dark radiation energy density, $\rho_{dr}\equiv f \rho_{r}$, evolving as $\rho_r = \rho_{r0}a^{-4}$ and represented by the coupling of the scalar field to the radiation. The last term can be interpreted as the constant dark energy contribution, $\Lambda = (1+f)V_0$, acting exactly like a cosmological constant term due to the coupling of the scalar field to the minimal value of the potential $V_0$. Notice that in the limit of a minimal coupling, $\xi\to 0$, the model reduces to a baryon, radiation and cosmological constant model, the last one coming from $V_0$. A more physical interpretation for the coupling between the scalar field to baryonic and radiation energy densities will be discussed at the end.

\section{Constraints from observational data}

Now will be presented the constraints of the free parameters of the model to observational data. Defining the present critical density $\rho_{c0} = \frac{3H_0^2}{8\pi G}$ and the density parameters:
\begin{equation}
    \Omega_b\equiv \frac{\rho_{b0}}{\rho_{c0}}\hspace{0.7cm} \Omega_r\equiv \frac{\rho_{r0}}{\rho_{c0}}\hspace{0.7cm} 
    \Omega_{dm}\equiv f\Omega_b = \frac{\alpha}{1-\alpha}\Omega_{b} 
    \hspace{0.7cm}
    \Omega_{dr}\equiv f\Omega_r = \frac{\alpha}{1-\alpha}\Omega_{r} \hspace{0.7cm}
    \Omega_\Lambda \equiv (1+f)\Omega^*_\Lambda = \frac{\Omega^*_\Lambda}{1-\alpha}\label{Om2}
\end{equation}
where $\Omega^*_\Lambda\equiv \frac{V_0}{\rho_{c0}}$, Eq. (\ref{H2f}) can be written in terms of the redshift $z$ as:
\begin{equation}
    \frac{H^2}{H_0^2}=\Omega_{b}(1+z)^3 + \Omega_{r}(1+z)^4 + \frac{\alpha}{1-\alpha}\Omega_{b}(1+z)^3 + \frac{\alpha}{1-\alpha}\Omega_{r}(1+z)^4 + \frac{\Omega^*_\Lambda}{1-\alpha}\,.\label{H2ff} 
\end{equation}
There are five independent free parameters, namely: $H_0,\,\Omega_b,\,\Omega_r,\,\Omega_\Lambda^*,\,\alpha$. The others can be obtained from (\ref{Om2}), as dark matter and dark radiation energy densities, beyond the cosmological constant like term. The Friedmann constraint for $z=0$ gives:
\begin{equation}
    1 = \Omega_b + \Omega_r + \Omega_b\frac{\alpha}{1-\alpha} + \Omega_r\frac{\alpha}{1-\alpha}  + \frac{\Omega_\Lambda^*}{1-\alpha}\,,\label{vinculo}
\end{equation}
and the number of free parameters are reduced to four. Since nothing is known about the dark radiation component and the present day value of the radiation is almost negligible, $\Omega_r \sim 5.39\times 10^{-5}$, it was chosen to fix the value of $\Omega_r$ and use a Gaussian prior up to $3\sigma$ c.l. to the baryonic density according to Planck 2018 survey \cite{Planck2018}, namely $\Omega_b = 0.0493 \pm 0.0026$. Since Eq. (\ref{vinculo}) can be used to eliminate one parameter from (\ref{H2ff}), one are left with a three free parameters model,  $H_0,\,\Omega_b,\,\alpha$, with $\Omega^*_\Lambda$ coming from (\ref{vinculo}).

\begin{figure}
\centering
\includegraphics[width=1.0\linewidth]{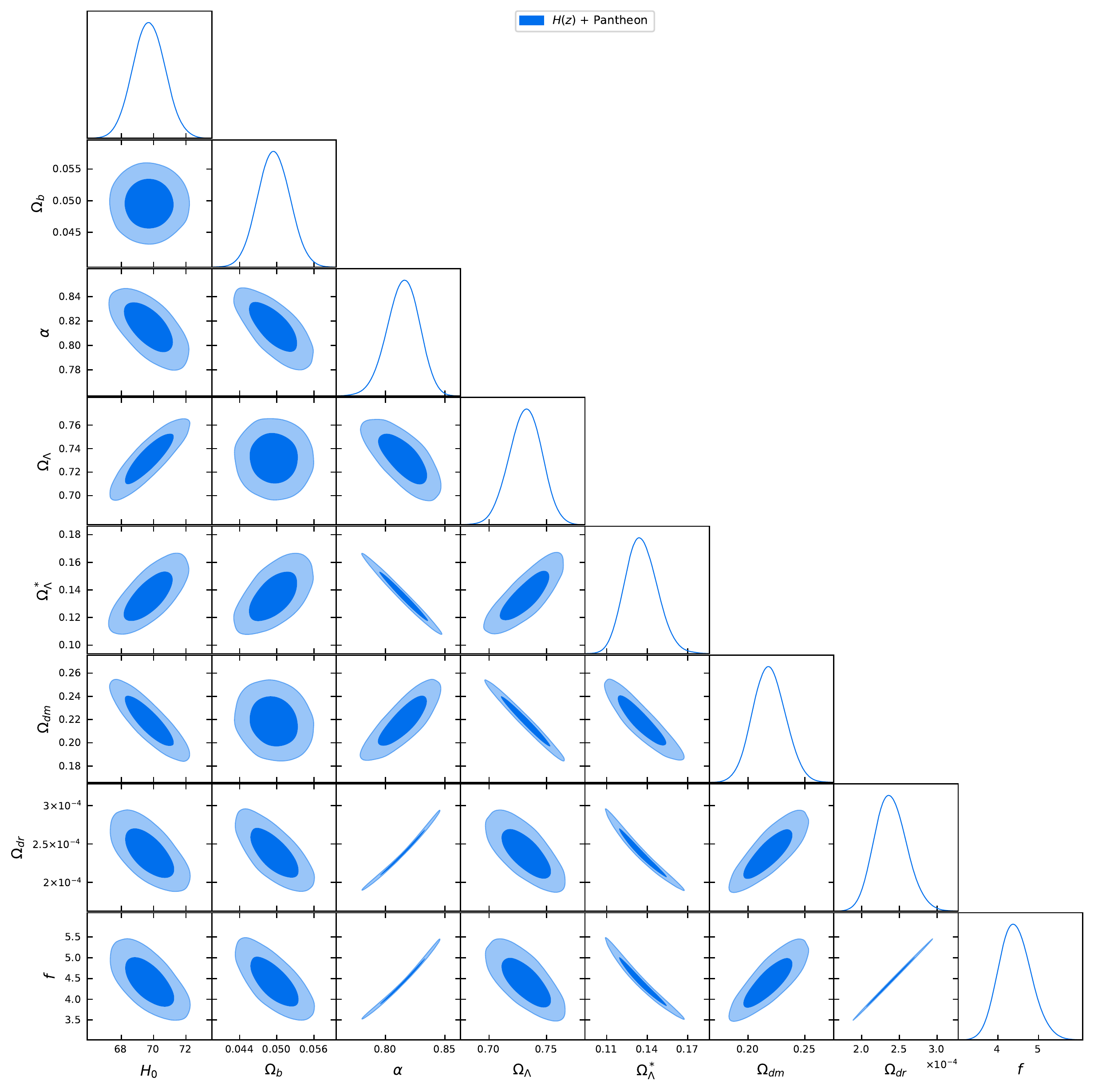}
\caption{Contours of the parameters at $1\sigma$ and $2\sigma$ c.l. with the combined analysis of $H(z)$ and SNeIa (Phanteon).}
\label{fig015}
\end{figure} 

The data used for the analysis were 51 $H(z)$ data from Maga\~na {\it et al.} \cite{Magana2018} and 1048 SNe Ia data from Pantheon compilation \cite{pantheon}. The constraints over the free parameters were done by sampling the likelihood $\mathcal{L} \propto e^{-\chi^2/2}$ through Monte Carlo Markov Chain analysis implemented in {\sffamily Python} language. The $\chi^2$ function for $H(z)$ data and for Pantheon data set are given by, respectively:
\begin{equation}
\chi^2_H = \sum_{i = 1}^{51}\frac{{\left[ H_{obs,i} - H(z_i,\mathbf{s})\right] }^{2}}{\sigma^{2}_{H_i,obs}} ,\hspace{1cm} \chi^2_{SN} = \left[\bm{m}_{obs}-{\bm m}(z,\bm{s})\right]^T\bm{C}^{-1}\left[\bm{m}_{obs}-{\bm m}(z,\bm{s})\right]
\label{chi2SN}
\end{equation}
where $\mathbf{s}=[H_0,\,\Omega_b,\,\alpha]$ is the parameter vector, $\bm{C}$, $\bm{m}_{obs}$ and $\bm{m}$ are covariance matrix, observed apparent magnitude vector and model apparent magnitude, respectively. We have assumed flat priors for $H_0$, $\alpha$ and a Gaussian prior for $\Omega_b$, within $3\sigma$ c.l..

The contours at $1\sigma$ and $2\sigma$ c.l. for the parameters $H_0,\,\Omega_b,\,\alpha$ and for the derived parameters $\Omega_{dm},\,\Omega_{dr},\,\Omega_\Lambda,\,\Omega_\Lambda^*,\, f$ are presented in Fig. \ref{fig015}. The mean values and 95\% c.l.  limits for each parameter are displayed in Table I.

\begin{table}
    \centering
   \begin{tabular} { l  c}

 Parameter &  95\% limits\\
\hline
{\boldmath$H_0            $} & $69.7^{+2.0}_{-2.0}        $\\

{\boldmath$\Omega_{b}     $} & $0.0495^{+0.0052}_{-0.0051}$\\

{\boldmath$\alpha         $} & $0.815^{+0.026}_{-0.028}   $\\

{$\Omega_\Lambda^*          $} & $0.136^{+0.025}_{-0.022}   $\\

$\Omega_\Lambda            $ & $0.732^{+0.028}_{-0.029}   $\\

$\Omega_{dm}               $ & $0.219^{+0.029}_{-0.028}   $\\

$\Omega_{dr}               $ & $0.000239^{+0.000045}_{-0.000040}$\\

$f                         $ & $4.43^{+0.84}_{-0.74}      $\\
\hline
\end{tabular}

    \begin{center}
         \caption{Mean value of the parameters and 95\% c.l. limits.}
    \end{center}
 \label{Tab01}
\end{table}

The first interesting result from Table I is that the condition $\alpha < 1$ is satisfied even at 95\% c.l., since that its mean value is $\alpha = 0.815^{+0.026}_{-0.028}$. Thus the convergence of the series (\ref{f}) is warranted. With this value for $\alpha$ one obtain the amplification factor $f=4.43^{+0.84}_{-0.74}$, which leads to a dark matter density parameter $\Omega_{dm}=0.219^{+0.029}_{-0.028}$, in good agreement to Planck 2018 results \cite{Planck2018} for the standard model. Additionally, the dark energy density parameter obtained is $\Omega_{\Lambda}=0.732^{+0.028}_{-0.029}$, which also agrees with the standard model result. The value for $H_0$ is also consistent to the Planck 2018 results. Notice that this simple model reproduces quite well all the background values for the main parameters of the concordance model.

Now the values of other parameters of the model can be estimated. With the value of $\alpha$ one obtain $\phi_0=\sqrt{\alpha/8\pi |\xi|}\,m_{pl} \simeq 0.0324\, m_{pl}/\sqrt{|\xi|}$. From the values of $\Omega^*_\Lambda$ and $H_0$ one obtain $V_0 = \Omega^*_\Lambda \rho_{c0} \simeq 5.34\times 10^{-50}$GeV$^4$. This value of $V_0$ is important to estimate the value of the physical mass $m$. For $A\simeq 6.0\times 10^{-3}m_{pl}$ and $\sigma \simeq 45.8m_{pl}$ the value of the physical mass is $m\approx 5\times 10^{-65} m_{pl}$ from (\ref{V0}). For $\xi\simeq 5\times 10^{-7}$ this case is in good agreement to inflation model ({\bf i}), since $\phi_f\simeq 46.1m_{pl} \gtrsim \phi_0 \approx \sigma \simeq 45.8 m_{pl}$. Additionally, these values furnish a DM/photon relation of about $1.5\times 10^{-29}$ if $\phi_* \simeq \phi_0$. An effective mass $m_e \simeq 2.2\times 10^{-6}m_{pl}$ is obtained, which leads to a correct density perturbation spectrum.

The inflationary case ({\bf ii}) is difficult to implement in this model, since that $\phi_f \simeq \sigma \sqrt{|\xi|} < \sigma$ for $|\xi| \ll 1$ and it is assumed $\phi_0 \simeq \sigma$ for the slowly varying regime.

For $\xi\simeq 10^{8}$ one obtain $\phi_f\simeq  \phi_0 \approx \sigma \simeq 3.2\times 10^{-6} m_{pl}$. This reproduces the inflationary case ({\bf iii}). With $A\simeq 2\times 10^{-6} m_{pl}$ the physical mass is $m\simeq 6.8\times 10^{-58} m_{pl}$ and $m_e\simeq 3.5\times 10^{-6}m_{pl} $. The DM/photon relation is satisfied only when the field value is $\sim m_{pl}$, long time before it reaches its final value.

\section{Concluding remarks}

A simple model of a scalar field non-minimally coupled to gravity was studied, under a potential involving a quadratic mass term plus a symmetry breaking one. It was showed that an inflationary phase is possible with reasonable  slow-roll parameters. Long time after inflation it is supposed that the field enters a slowly varying phase at a nearly constant value $\phi_0$. The nonminimal coupling $\xi$ between the scalar field and the Ricci scalar is responsible for an effective coupling of the scalar field to the standard baryonic and radiation energy density part o the Lagrangian, producing a dark matter and dark radiation contribution. Also an effective cosmological constant term naturally appears, proportional to the minimum of the potential.

The constraints of the free parameters $H_0, \,\Omega_b, \alpha$ with observational data furnish values for $\Omega_{dm}$ and $\Omega_{\Lambda}$ that are in good agreement with the values of Planck 2018 results for standard model. Additionally a term of dark radiation $\Omega_{dr}$ appears naturally. The constant $\alpha$, related to the nonminimal coupling and the present day value of the scalar field through $\alpha = \kappa^2\xi\phi_0^2 = 8\pi G \xi \phi_0^2$, produces a kind of amplifier function $f = \alpha/(1-\alpha)$. Such function is responsible to relate the baryonic density parameter $\Omega_b$ to the dark matter density parameter, $\Omega_{dm}=f\Omega_{b}$. The same occurs with the radiation density parameter, $\Omega_{r}$, producing a kind of dark radiation density,  $\Omega_{dr}=f \Omega_{r}$. Notice that the mean value obtained for $f$, namely $f=4.43^{+0.84}_{-0.74}$, is exactly the one necessary to `amplify' the effect of the baryonic energy density and produces the expected dark matter energy density, $\Omega_{dm}=0.219^{+0.029}_{-0.028}$. The same occurs for the dark energy like term, namely $\Omega_\Lambda=(1+f)\Omega^*_\Lambda =0.732^{+0.028}_{-0.029}$. The other parameters of the model, as the physical mass of the scalar field and mass scales of the potential, correctly reproduces at least one inflationary scenery, with $\xi \ll 1$. Additionally, the DM/photon ratio agrees to expected and the effective mass found is also the expected ones to reproduces correct density perturbations at the end of inflation.

The physical meaning of the coupling must be better explored. For experiments that are sensitive just to standard model sector, what the experiment ``sees" is the net value of $\rho_b$ and $\rho_r$ at the level of the Friedmann equation (\ref{H2a}). For this case it is the gravitational constant $G$ that must be rescaled, which does not alter electromagnetic based experiments.  In order to test the dark matter or dark radiation sector the experiment must be sensitive to the scalar field $\phi$, then the experiment ``sees" the Friedmann equation (\ref{H2f}), where the coupling of the scalar field to radiation and baryonic sector is evident. For this case the gravitational constant is not altered, however the ``amplification" of the gravitational effects of the baryonic and radiation content comes through $f$. Notice that the minimum value of the potential is also amplified, acting as a dard energy component.

As a last analysis from a quantum point of view, if the field is considered as a wave function, the quantity $\alpha$ can also be interpreted as the present expected value of the coupling of $\xi$ to the gravitational constant $G$ through $\alpha = \langle \phi_0 | 8\pi G \xi |\phi_0 \rangle$. Thus, the amplification effect $f$ can be interpreted as the contribution of the wave function of the scalar field on the gravitational coupling. 

In conclusion, it is important to emphasize that this work is just a kind of ``toy model", which presents another alternative to interpret the dark components of the universe, through a new mechanism of coupling the scalar field with the standard matter. The dynamical equations must be better studied and also the same kind of coupling in other metrics, in order to better clarify the mechanism at different systems.

\begin{acknowledgements}
The author thank Jos\'e F. Jesus for valuable discussions on observational data analysis. This work was supported by the CNPq - Conselho Nacional de Desenvolvimento Cient\'ifico e Tecnol\'ogico, Brazilian research agency, grant number 303583/2018-5.  
\end{acknowledgements}


\end{document}